\documentstyle[prd,aps,epsf,floats,amsfonts,amssymb,amsmath]{revtex}

\begin{document}
 \draft
 \title{Effects of CPT and Lorentz Invariance Violation on Pulsar Kicks}
% \title{Constraints on Parameters Breaking the Lorentz Invariance from Pulsar Kicks}
%  \title{Pulsar Kick Induced by Lorentz Invariance Breakdown}
%\author{xxx}
%\address{yyy}
\author{G. Lambiase$^{a,b}$}
\address{$^a$Dipartimento di Fisica "E.R. Caianiello"
 Universit\'a di Salerno, 84081 Baronissi (Sa), Italy.}
  \address{$^b$INFN - Gruppo Collegato di Salerno, Italy.}
\date{\today}
\maketitle
\begin{abstract}
The breakdown of Lorentz's and CPT invariance, as described by the
Extension of the Standard Model, gives rise to a modification of
the dispersion relation of particles. Consequences of such a
modification are reviewed in the framework of pulsar kicks induced
by neutrino oscillations (active-sterile conversion). A peculiar
feature of the modified energy-momentum relations is the
occurrence of terms of the form $\delta {\bbox \Pi}\cdot {\bf
{\hat p}}$, where $\delta {\bbox \Pi}$ accounts for the difference
of spatial components of flavor depending coefficients which lead
to the departure of the Lorentz symmetry, and ${\bf {\hat p}}={\bf
p}/p$, being ${\bf p}$ the neutrino momentum. Owing to the
relative orientation of ${\bf p}$ with respect to $\delta {\bbox
\Pi}$, the {\it coupling} $\delta {\bbox \Pi}\cdot {\bf {\hat p}}$
may induce the mechanism to generate the observed pulsar
velocities. Topics related to the velocity distribution of pulsars
are also discussed.
\end{abstract}
% \Keyword(s){Lorentz and Poincare invariance, Charge conjugation-ParityTime reversal,Pulsar, Neutrino mass and mixing}
%\vskip2pc]

%\section{Introduction}
%\setcounter{equation}{0}

The studies of a possible breakdown of the fundamental symmetries
in physics represent a very active research area. As suggested by
Kosteleck\'y and Samuel, String/M theory provides a scenario in
which a departure of the Lorentz invariance might manifest
\cite{kostelecky}. Recently, these investigations have been
reconsidered in the context of $D$-branes \cite{ellis}, Loop
Quantum Gravity \cite{alfaroPRL,alfaroPRD,loop}, Non-Commutative
Geometry \cite{NCG}, and through the spacetime variation of
fundamental coupling constants \cite{bertolami}. For modern tests
on Lorentz invariance, see \cite{mattingly}

According to Ref. \cite{kostelecky}, Lorentz's invariance
violation (due to non trivial solution of (open) string field
theory) follows from the observation that the vacuum solution of
the theory could spontaneously violate the Lorentz and CPT
invariance, even though such symmetries are satisfied by the
underlying theory. The breakdown of these fundamental symmetries
occurs in the Extension of the Standard Model, and only operators
of mass with dimension four or less \cite{calladay,lehnert} are
involved in order that the standard model power-counting
renormalizability is preserved. In a recent work by Kosteleck\'y
and Mewes \cite{mewesneutrino}, it has been studied the general
formalism for violations of Lorentz and CPT symmetry in the
neutrino sector. The generalized equation of motion for free
fermions (neutrinos in our case) is given by (we shall use natural
units $c=1=\hbar$)
\begin{equation}\label{EqOfMotion}
  \left(i\,\Gamma^\nu_{AB} \partial_\nu - M_{AB}\right)\psi_B=0\,,
  \end{equation}
where the spinor $\psi_B$ contains all the fields and their
conjugates, the indices $A$ and $B$ range over all $2N$
possibility $\{f,f^C\}$, being $f=e, \mu, \tau,\ldots$ the
neutrino flavors and $f_A^C={\cal C}_{AB}f_B$. ${\cal C}$ is the
symmetric matrix with non zero components ${\cal C}_{ff^C}=1$.
$\Gamma^\nu_{AB}$ and $M_{AB}$ are $4\times 4$ matrices in the
spinor space, and can be decomposed using the basis of $\gamma$
matrices
\begin{equation}\label{Gamma}
  \Gamma^\nu_{AB}=\gamma^\nu_{AB}+c^{\mu\nu}_{AB}\gamma_\mu+
  d^{\mu\nu}_{AB}\gamma_5\gamma_\mu
  +e^\nu_{AB}+if^\nu_{AB}\gamma_5+
  \frac{1}{2}g^{\lambda\mu\nu}_{AB}\sigma_{\lambda\mu}\,,
\end{equation}
and
\begin{equation}\label{M}
  M_{AB}=m_{AB}+im_{5AB}\gamma_5+a_{\mu AB}\gamma^\mu+
  b_{\mu AB}\gamma_5\gamma^\mu+\frac{1}{2}\,
  H^{\mu\nu}_{AB} \sigma_{\mu\nu}\,.
\end{equation}
The coefficients $a_{\mu AB}$, $b_{\mu AB}, c^{\mu\nu}_{AB},
\ldots$ are constants and in general they are flavor depending,
and $\sigma^{\mu\nu}=\frac{1}{4}[\gamma^\mu, \gamma^\nu]$.
$c^{\mu\nu}_{AB}$, $d^{\mu\nu}_{AB}$, and $H^{\mu\nu}_{AB}$
preserve the CPT invariance, while $a_{\mu AB}$, $b_{\mu AB}$,
$e_{\mu AB}$, $f_{\mu AB}$ violate CPT and Lorentz invariance.
Finally $m$ and $m_5$ are Lorentz and CPT conserving.

The time evolution of neutrinos is governed by the effective
Hamiltonian
\begin{eqnarray}\label{heff}
  (H_{eff})_{ab} &=& p\,\delta_{ab} \left(\begin{array}{cc}
                           1 & 0 \\
                              0 & 1 \end{array}\right)+
                              \frac{1}{2p}\left(\begin{array}{cc}
                           ({\tilde m}^2)_{ab} & 0 \\
                             0 & ({\tilde m}^2)_{ab}^*
                             \end{array}\right) +  \\
                & & + \frac{1}{p} \left(\begin{array}{cc}
                           [(a_L)^\mu p_\mu-(c_L)^{\mu\nu}p_\mu p_\nu]_{ab} & -i\sqrt{2}\, p_\mu (\epsilon_+)_\nu
                                                    [(g^{\mu\nu\sigma}p_\sigma-H^{\mu\nu}){\cal C}]_{ab}
                                                   \vspace{0.3cm} \\
                i\sqrt{2}\, p_\mu (\epsilon_+)^*_\nu [(g^{\mu\nu\sigma}p_\sigma+H^{\mu\nu}){\cal C}]_{ab}^* &
                [-(a_L)^\mu p_\mu-(c_L)^{\mu\nu}p_\mu p_\nu]^*_{ab}
                             \end{array}\right)\,, \nonumber
\end{eqnarray}
where $(c_L)^{\mu\nu}_{ab}=(c+d)_{ab}^{\mu\nu}$ and
$(a_L)^{\mu}_{ab}=(a+b)_{ab}^{\mu}$, the vector
$(\epsilon_+)^\nu=\frac{1}{\sqrt{}2}(0, {\bbox \epsilon}_1+i{\bbox
\epsilon}_2)$ is related to the helicity of neutrinos (${\bbox
\epsilon}_1$ and ${\bbox \epsilon}_1$ are two real vectors), while
$({\tilde m}^2)_{ab}$ is related to the usual neutrino masses.
Details of properties of the coefficients entering in the
effective Hamiltonian can be found in \cite{mewesneutrino}. It is
worth to note that Eq. (\ref{heff}) involves only the coefficients
$a$, $c$, $g$ and $H$, since the other coefficients breaking the
Lorentz invariance can be removed, at the leading order, with an
appropriate redefinition of fields.

Bounds on parameters entering in the Extension of the Standard
Model which involve neutrino oscillations have been discussed in
\cite{mewesneutrino,KostLSND,KostPRD70,MMewes,gaetanoPLB}.
Stringent bounds on the coefficients $(a_L)^\mu$ and
$(c_L)^{\mu\nu}$ have been obtained in
\cite{mewesneutrino,KostLSND,KostPRD70} by analyzing various type
of experiments (solar neutrino experiments [SK, SNO]; atmospheric
neutrinos experiments [SK, KamLAND, LSND, K2K]; reactor
experiments [CHOOZ, Palo Verde]; short-baseline experiments
[CHORUS, NOMAD, KARMEN]; long-baseline experiments [ICARUS, MINOS,
OPERA]). Estimations of the attainable sensitivities to the
dimensionless ($\delta c_L=c_L^f-c_L^{f'}$) and dimension-one
($\delta a_L=a_L^f-a_L^{f'}$) coefficients are
\cite{mewesneutrino}
\begin{equation}\label{cLaL}
  10^{-26}\lesssim \delta c_L^{\mu\nu} \lesssim 10^{-17}\,, \qquad
  10^{-19}\mbox{eV} \lesssim \delta a_L^\mu \lesssim
  10^{-9}\mbox{eV}\,.
\end{equation}
%As discussed in \cite{mewesneutrino}, the expected values for
%$\delta c_L^{\mu\nu}$ and $\delta a_L^\mu$ are around $10^{-20}$
%and $10^{-11}$eV, respectively.

The aim of this paper is to investigate the consequences of the
Lorentz invariance breakdown in relation to pulsar kicks induced
by neutrino oscillations in matter. The origin of the pulsar
velocity is till now an open issue of the modern astrophysics. As
follows from the observations, pulsars have a very high velocity,
in comparison with the surrounding stars, which may varies from
$450\pm 90$Km/sec up to values greater than 1000Km/sec
\cite{Lyne,arzoumanian}. This suggests that nascent pulsars
undergo to some kind of kick. After the supernova collapse of a
massive star, neutrinos carry away almost all ($99\%$) the
gravitational binding energy ($3\times 10^{53}$erg). The momentum
taken by them is about $10^{43}$gr cm/sec. A fractional anisotropy
of the order $\sim 1\%$ of the outgoing neutrino momenta would
suffice to account for the neutron star recoil of 300Km/sec. Many
mechanisms have been proposed to solve such a issue, but a
definitive solution is still laking.

An elegant and interesting mechanism to generate the pulsar
velocity, which involves the neutrino oscillation physics, has
been proposed by Kusenko and Segr\'e \cite{kusenkoPRL}. Under
suitable conditions, a resonant oscillation $\nu_e\to \nu_{\mu,
\tau}$ may occur in the region between the corresponding ($\nu_e$
and $\nu_{\mu, \tau}$) neutrinospheres. The $\nu_e$ are trapped by
the medium (due to neutral and charged interactions) but neutrinos
$\nu_{\mu, \tau}$ generate via oscillations can escape from the
protostar being outside of their neutrinosphere. Thus, the surface
of the resonance acts as an effective muon/tau neutrinosphere. In
the presence of a magnetic field ${\bf B}$, the surface of
resonance is distorted with the ensuing that the energy flux turns
out to be generated anisotropically (neutrinos generated in
regions with different temperatures are emitted with different
energies). Indeed, in a magnetized medium, the resonance condition
in matter is modified by a term $\sim {\bf B}\cdot {\bf p}$, where
${\bf p}$ the neutrino momentum. The relative orientation of the
neutrino momentum with respect to the magnetic field generate the
asymmetry in the neutrino emission. The case in which
active-sterile neutrinos are involved has been studied in
\cite{kusenkoPLB}. Papers dealing with the problem of the origin
of pulsar velocities, can be found in Refs.
\cite{alex,other,fuller,zanella,zanella0,casini,nardi,burrows,cuesta,all,gaetanoPRL}.

The effective Hamiltonian (\ref{heff}) suggests a further
mechanism to generate pulsar kicks. As we will see, terms breaking
the Lorentz invariance give rise to a coupling of the form $\sim
\delta {\bbox \Pi}\cdot {\bf {\hat p}}$, where $\delta {\bbox
\Pi}$ is the difference of spatial components $c_L^{0i}$, $a_L^i$
referred to different flavors, and ${\bf {\hat p}}={\bf p}/p$,
where $p=|{\bf p}|$. The relative orientation of neutrino momenta
with respect to $\delta {\bbox \Pi}$ determines an asymmetry in
the neutrino emission, hence can generate the pulsar kicks. This
effect might provide a signature of a possible violation of the
Lorentz invariance.

%\section{Pulsar Kicks Induced by Lorentz Invariance Breakdown}
%\setcounter{equation}{0}

As observed in \cite{kusenkoPLB}, oscillations of active neutrinos
could be a plausible explanation of the observed pulsar velocities
if the resonant conversion $\nu_{\tau, \mu}\leftrightarrow \nu_e$
occurs between two different neutrinospheres. However, the
resonant transition between two neutrinoshperes leads to neutrino
masses which do not agree with the present limits on the masses of
standard electroweak neutrinos. These limits do not apply to
sterile neutrinos that may have only a small mixing angle with the
ordinary neutrinos \cite{kusenkoPLB}. Thus we shall confine
ourselves to active-sterile conversion of neutrinos.

Eq. (\ref{heff}) implies that the energy of neutrinos with a given
flavor is (we shall consider only the diagonal terms)
\begin{equation}\label{disp-helicity}
  E\simeq p+\frac{m^2}{2p} +\Omega\,,
\end{equation}
with
\begin{equation}\label{Omega0}
  \Omega=-\frac{c^{\mu\nu}_Lp_\mu p_\nu}{p}+\frac{a_{L\,\mu}
  p^\mu}{p}\,.
\end{equation}
For our purpose, it is convenient to rewrite Eq. (\ref{Omega0}) in
the form
\begin{equation}\label{Omega01}
  \Omega\approx \Pi_0+{\bbox \Pi}\cdot {\hat {\bf p}}\,,
\end{equation}
where
\begin{eqnarray}\label{Gamma0}
 \Pi_0 &=& -c^{00}_L \,p+\frac{c^{ij}_L p_ip_j}{p}+a_{L\,0}\,, \\
 {\bbox \Pi} & = & -2\,p\,{\bf c}_L+{\bf a}_L\,,
 \label{Gammai}
\end{eqnarray}
and the relations $p^0\simeq p$ and $c^i=c^{0i}$ have been used.

To estimate the anisotropy of the outgoing neutrinos, one needs to
evaluate the energy flux ${\bf F}_s$ emitted by nascent stars.
According to Ref. \cite{zanella}, the asymmetry of the neutrino
momentum is
\begin{equation}\label{asymmetrygeneral}
  \frac{\Delta p}{p}\cong \frac{1}{3}
  \frac{\int_0^\pi {\bf F}_s(\vartheta)\cdot \delta {\bf {\hat \Pi}} \, da}
  {\int_0^\pi {\bf F}_s(\vartheta)\cdot {\bf {\hat n}} \, da }
  \sim -\frac{1}{9}\frac{\varrho}{r_{res}}\,.
\end{equation}
The factor 1/3 takes into account of the fact that only one
neutrinos species is responsible for the anisotropy (thus it
carries out only 1/3 of the total energy). ${\bf F}_s$ is the
outgoing neutrino flux through the element area $da$ of the
emission surface. $\delta {\bf {\hat \Pi}}={\bbox \Pi}/|{\bbox
\Pi}|$ is the direction of the vector $\delta {\bbox \Pi} = {\bbox
\Pi}^{(\nu_f)}-{\bbox \Pi}^{(\nu_s)}$, whereas ${\bf {\hat n}}$ is
the unity vector orthogonal to the element area $da$. The
parameter $\varrho$ is the radial deformation of the effective
surface of resonance generated by the term breaking the Lorentz
invariance ${\bbox \Pi}$. It shifts the resonance point $r_{res}$
to $r(\phi)=r_{res}+\varrho \cos\phi$, with $\varrho\ll r_{res}$
and $\cos \phi= \delta {\bbox {\hat \Pi}} \cdot {\bf {\hat p}}$.
The resonance point $r_{res}$ is determined by the usual resonance
condition (MSW effect \cite{MSW})
 \begin{equation}\label{res}
 2\delta c_2-V_{\nu_f}(r_{res})=0\,,
 \end{equation}
where
 \begin{equation}\label{oscparam}
 \delta=\frac{\Delta m^2}{4p}\,, \qquad c_2=\cos 2\theta\,.
 \end{equation}
$\Delta m^2=m_2^2-m_1^2$ is the mass squared difference (in the
notation of (\ref{oscparam}), we shall also indicate $\sin
2\theta=s_2$). In (\ref{res}), the potential $V_{\nu_f}$ is
defined as follows
\begin{eqnarray}\label{Vnue}
 V_{\nu_e} &=& -V_{{\bar \nu}_e}=V_0(3Y_e-1+4Y_{\nu_e})\,, \\
 V_{\nu_{\mu,\tau}} &=& -V_{{\bar
 \nu}_{\mu,\tau}}=V_0(Y_e-1+2Y_{\nu_e})\,, \label{Vnuf}
\end{eqnarray}
where $Y_e$ ($Y_{\nu_e}$) represents the ratio between the number
density of electrons (neutrinos), and
\begin{equation}\label{V0}
  V_0=\frac{G_F\rho}{\sqrt{2}m_n}=\frac{\rho}{10^{14}\mbox{gr/cm}^3}\,\,
  3.8\,\mbox{eV}\,.
\end{equation}
$m_n=938$MeV is the nucleon mass and $\rho$ the matter density.
For sterile neutrinos one has $V_{\nu_s}=0$.

The equation of evolution describing the conversion between two
neutrino flavors (we consider the conversion $\nu_f\leftrightarrow
\nu_s$, $f=e, \mu, \tau$) is
 \begin{equation}\label{11}
i\frac{d}{d r}\left(\begin{array}{c}
                           \nu_{f} \\
                              \nu_s \end{array}\right)={\cal H}\left(\begin{array}{c}
                           \nu_{f} \\
                             \nu_s\end{array}\right)\,,
 \end{equation}
where the matrix ${\cal H}$ is the effective Hamiltonian defined
as
\begin{equation}\label{12}
{\cal H}=\left[\begin{array}{cc}
 V_{\nu_f}-c_2\delta + {\bbox\Pi}^{(\nu_f)}\cdot {\bf {\hat p}} & s_2\delta \vspace{0.05in} \\
 s_2\delta & c_2\delta + {\bbox \Pi}^{(\nu_s)}\cdot {\bf {\hat p}}
\end{array}\right]\,,
\end{equation}
up to terms proportional to identity matrix\footnote{A comment is
in order. The Lorentz invariance violation implies that the
propagation of neutrinos depends, in general, on their flavors
through the parameters $c^{\mu\nu}_L$, $a_{L\,\mu}$ (for the
moment, we shall neglect neutrino masses). The velocity is
 \[
 v_f=\frac{dE}{dp}\approx 1-\frac{d\Omega}{dp}\,.
 \]
Different neutrino species may have different maximum attainable
velocities \cite{coleman,leung}. This occurs if neutrino flavor
eigenstates are distinct from neutrino velocity eigenstates, being
the two eigenstates related linearly by the mixing angle
$\theta_v$. Thus, the diagonal terms ${\bbox \Pi}^{(\nu_f,
\nu_s)}\cdot {\bf {\hat p}}$ in (\ref{12}) should be replaced by
${\bbox \Pi}^{(\nu_f, \nu_s)}\cdot {\bf {\hat p}}\cos 2\theta_v$,
whereas the term ${\bbox \Pi}^{(\nu_f, \nu_s)}\cdot {\bf {\hat
p}}\sin 2\theta_v$ should appear in the off-diagonal. For
simplicity we have taken $\theta_v=0$.}.

As follows from (\ref{12}), the resonance condition reads
\begin{equation}\label{rescondLIV}
  2\delta c_2=V_{\nu_f}+\delta {\bbox \Pi}\cdot {\bf {\hat p}}
  +\delta \Pi_{0} \,.
\end{equation}
where
 \begin{eqnarray}\label{deltaGamma}
 \delta {\bbox \Pi} \cdot {\bf {\hat p}} &=&  (\delta {\bf a}-2p\delta {\bf c})\cdot {\bf {\hat
  p}} \\
  & = & |\delta {\bf a}-2p\delta {\bf c}|\cos\phi\,,
  \nonumber
 \end{eqnarray}
and $\delta {\bf c}= {\bf c}^{(\nu_f)}-{\bf c}^{(\nu_s)}$, $\delta
{\bf a}= {\bf a}^{(\nu_f)}-{\bf a}^{(\nu_s)}$. A similar
expression holds for $\delta \Pi_{0}$. Notice that $\delta
\Pi_0({\bf p})=\delta \Pi_0(-{\bf p})$ and, for typical values of
$\rho\sim (10^{11}-10^{14})$gr/cm$^3$ in the protostar, $\delta
\Pi_{0} \ll 2\delta c_2$ according to constraints (\ref{cLaL}).
Moreover, since $\delta {\bbox \Pi}\cdot {\bf {\hat p}}$ changes
sign under the transformation ${\bf p}\to -{\bf p}$, it may deform
the resonance surface, a condition necessary in order that the
asymmetry of the outgoing neutrino momenta may occur.

Let us now evaluate $\varrho$. Expanding the terms in
(\ref{rescondLIV}) about to $r=r_{res}+\varrho \cos \phi$, with
$p\to p+\delta_p$, $V_{\nu_f}\to V_{\nu_f}+ \delta_{V_{\nu_f}}$
\cite{zanella}, in which
 \begin{eqnarray}
 \delta_p&=&\frac{d\ln p}{dr}\,p\,\varrho\,\cos \phi=h_p^{-1}p\,\varrho\cos \phi\,,
 \label{deltap} \\
  \delta_{V_{\nu_f}}&=&\frac{d\ln V_{\nu_f}}{dr}\,V_{\nu_f}\,\varrho\,\cos\phi=
  h_{V_{\nu_f}}^{-1}\,V_{\nu_f}\,\varrho\cos \phi\,,
  \label{deltaV}
  \end{eqnarray}
and using the resonance condition (\ref{res}), one infers (all
quantities are evaluated at the resonance)
\begin{equation}\label{deltaBOmega}
  \varrho \approx
  -\frac{\Sigma}{V_{\nu_f}}  \frac{1}{h_p^{-1}+h_{V_{\nu_f}}^{-1}}\,,
\end{equation}
where
\begin{equation}\label{Sigma}
  \Sigma\equiv 2p|\delta {\bf c}_L|+|\delta {\bf a}_L|\,.
 \end{equation}
Inserting Eq. (\ref{deltaBOmega}) into Eq.
(\ref{asymmetrygeneral}) one gets
\begin{equation}\label{deltap1}
  \frac{\Delta p}{p}=\frac{1}{9}\frac{\Sigma}{V_{\nu_f}}\frac{1}{r_{res}(h_p^{-1}+h_{V_{\nu_f}}^{-1})}\,,
\end{equation}
which implies that the fractional asymmetry $\Delta p/p$ is $\sim
1\%$ provided
\begin{equation}\label{Sigma1}
  \Sigma\sim
  0.09\,V_{\nu_f}r_{res}(h_p^{-1}+h_{V_{\nu_f}}^{-1})\,.
\end{equation}
To compute $h_p^{-1}+h_{V_{\nu_f}}^{-1}$ one has to specify a
model for the protostar. To this aim, we assume that the inner
core of a protostar is consistently described by a polytropic gas
of relativistic nucleon with adiabatic index $\Gamma=4/3$
\cite{shapiro}. The pressure $P$ and matter density $\rho$ are
related by \cite{shapiro,zanella}
\begin{equation}\label{pressure}
  P=K\rho^\Gamma\,,
\end{equation}
where $K=T_c/m_n\rho_c^{1/3}\simeq 5.6\times 10^{-5}$MeV$^{-4/3}$.
Here $T_c=40$MeV and $\rho_c=2\times 10^{14}$gr/cm$^3$ are the
temperature and the matter density of the core, respectively.

The matter density $\rho(r)$ can be expressed in the form
\cite{zanella} (see Appendix)
\begin{equation}\label{rhoG}
 \rho^{\Gamma -1}(x)=\rho_c^{\Gamma-1}[a' x^2+b' x+c']\,,
\end{equation}
where
\begin{equation}\label{abc}
  x=\frac{r_c}{r}\,,\quad a'=(1-\mu)\lambda_\Gamma\,,\quad
  b'=(2\mu-1)\lambda_\Gamma\,, \quad c'=1-\mu\lambda_\Gamma\,,
\end{equation}
and $r_c=10$km is the core radius. The parameter $\mu$ is
determined by setting $\rho(R_s)=0$ ($R_s$ the radius of the star)
\begin{equation}\label{mu}
  \mu=\left[\frac{R_s}{\lambda_\Gamma(R_s-r_c)}-\frac{r_c}{R_s}\right]
  \frac{R_s}{R_s-r_c}\,.
\end{equation}
$\lambda_\Gamma$ in (\ref{mu}) and (\ref{abc}) is given by
\begin{equation}\label{lambdaG}
  \lambda_\Gamma=\frac{G_N M_c}{r_c\rho_c^{\Gamma-1}}\frac{\Gamma-1}{K\Gamma}
  \simeq 0.29\, \frac{2G_N
  M_c}{\mbox{km}}\frac{10\mbox{km}}{r_c}\frac{40\mbox{MeV}}{T_c}\,,
\end{equation}
where $M_c\simeq M_\odot$ is the mass of the core ($M_\odot$ is
the solar mass). The temperature profile $T(r)$ is related to
matter density $\rho(r)$ through the relation \cite{zanella} (see
Appendix)
\begin{equation}\label{Tprofile}
  \frac{dT^2}{dr}=-\frac{9\kappa L_c}{\pi r^2}\, \rho\,,
\end{equation}
where $L_c$ is the core luminosity, $L_c\sim 9.5\times
10^{51}$erg/sec, and $\kappa=5.6\times
10^{-9}$cm$^4$/erg$^3$sec$^2$ $\sim 6.2\times 10^{-56}$eV$^{-5}$.
Eq. (\ref{Tprofile}) can be immediately integrate by using
(\ref{rhoG})
\begin{equation}\label{T(r)}
  T(r)=T_c\, \sqrt{2\lambda_c[\chi(x)-\chi(1)]+1}\,,
\end{equation}
where
\begin{equation}\label{chi}
  \chi(x)=c^{\prime\,3}x+\frac{3}{2}\, b^{\prime}\,c^{\prime 2}x^2+c^{\prime}(a^{\prime}\,c^{\prime}+
  b^{\prime\,2})\,x^3
  +\frac{b^{\prime}}{4}\,(6\,a^{\prime}\,c^{\prime}+b^{\prime\,2})\,x^4+
  \frac{3\,a^{\prime}}{5}\,(a^{\prime}\,c^{\prime}+b^{\prime\,2})\,x^5
  +\frac{b^{\prime}\,a^{\prime \, 2}}{2}\,x^6+\frac{a^{\prime \,3}}{7}\,x^7\,,
\end{equation}
and
\begin{equation}\label{lambdac}
  \lambda_c=\frac{9}{2\pi}\frac{\kappa \, L_c \, \rho_c}{T_c^2 r_c}
  \sim 1.95\, \frac{\rho_c}{10^{14}\mbox{gr/cm}^3}\,\frac{10\mbox{km}}{r_c}
  \,\left(\frac{40\mbox{MeV}}{T_c}\right)^2\,.
\end{equation}
Since $p\sim T$ (it is assumed the thermal equilibrium between
neutrinos and the medium, so that the average energy of the
emitted neutrino is proportional to the temperature at the
emission point \cite{zanella}) and $V_{eff}\sim \rho$, one can
rewrite the inverse characteristic lengths $h_p^{-1}$ and
$h_{V_{\nu_f}}^{-1}$ as $h_p^{-1}\equiv h_T^{-1}$ and
$h_{V_{\nu_f}}^{-1}\equiv h_\rho^{-1}$ \cite{zanella}. Eqs.
(\ref{rhoG}) and (\ref{T(r)}) imply (at the resonance)
\begin{eqnarray}
 h_T^{-1} &=& \frac{d\ln T}{dr} = -\lambda_c\, \frac{\rho(r_{res})}{\rho_c}\,
        \left(\frac{T_c}{T(r_{res})}\right)^2\,\frac{x_{res}}{r_{res}}\,,
        \label{hT} \\
 h_\rho^{-1} &=& \frac{d\ln \rho}{dr} = -3\, \left(\frac{\rho_c}{\rho(r_{res})}\right)^{1/3}\,
        \left(2\,a\, x_{res}+b\right)\,\frac{x_{res}}{r_{res}}\,,  \label{hV}
\end{eqnarray}
so that Eq. (\ref{Sigma1}) reads
\begin{equation}\label{Sigma2}
  \Sigma\sim 0.09\, V_{\nu_f}\, \lambda_\Gamma \,x_{res}\, \eta\,,
\end{equation}
where
\begin{equation}\label{eta}
  \eta\equiv \varepsilon^2\,
  \frac{\lambda_c}{\lambda_\Gamma}+3(2\mu-1)-6(\mu-1)x_{res}\,,
\end{equation}
and the parameter $\varepsilon=T_c/T(r_{res})$ has been
introduced. Eq. (\ref{eta}) gives a constraint on possible values
of $\mu$, $\varepsilon$ and $x_{res}$.

The Lorentz and CPT symmetry breakdown is relevant (on pulsar
kicks) for resonances occurring at $\rho(r_{res})=\rho_c$. In such
a case, Eq. (\ref{Vnuf}) reduces to $V_{\nu_f}\simeq 0.7V_0$ (we
use $Y_e\sim Y_{\nu}\sim {\cal O}(10^{-1}))$. The mass of the
sterile neutrino ($m_{\nu_s}\gg m_{\nu_f}$) is derived through Eq.
(\ref{res}) for small mixing angle and $p\sim 20$MeV. One gets
$m_{\nu_s}\sim$ few keV, according to Ref. \cite{fuller}. Eq.
(\ref{Sigma2}) then becomes
\begin{equation}\label{Sigma3}
  \Sigma\sim 2\times 10^{-2}x_{res}\,\eta \,\mbox{eV}\,.
\end{equation}
Eqs. (\ref{Sigma}) and (\ref{Sigma3}), and $x_{res}\sim {\cal
O}(1)$ (as we shall determine below) yield
\begin{equation}\label{coeff-c-Core}
  |\delta {\bf c}_L| \lesssim 4\times  10^{-10}\eta\,, \qquad
  |\delta{\bf a}_L| \lesssim  1.6 \times 10^{-2}\eta\, \mbox{eV}\,.
\end{equation}
The constraints (\ref{cLaL}) provide $\eta\lesssim 10^{-7}$.

\begin{figure}
\centering \leavevmode \epsfxsize=7cm \epsfysize=5cm
\epsffile{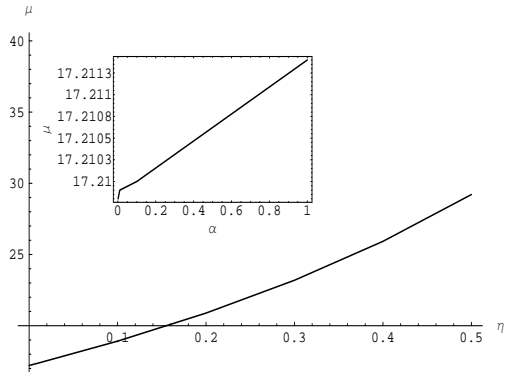} \caption{The parameter $\mu$, solution of Eq.
(\ref{eqpermu}), vs the parameter $\eta$. $\alpha$ is defined as
$\alpha=\eta/10^{-4}$, and, in the small panel it varies from 1 to
$10^{-3}$.} \label{fig}
\end{figure}

Eq. (\ref{rhoG}) admits the solutions $x_{res}=1$ and $x_{res}=\mu
/(\mu-1)$. The first solution is not compatible with (\ref{eta})
since it implies $\eta=\varepsilon^2\lambda_c/\lambda_\Gamma+3$,
so that no real solution there exists for $\varepsilon$. The
second one inserted into Eq. (\ref{eta}) yields
 \[
 \varepsilon = \sqrt{\frac{\lambda_\Gamma(3+\eta)}{\lambda_c}}\simeq 0.82\,.
 \]
Finally, the parameter $\mu$ is determined via Eq. (\ref{T(r)})
\begin{equation}\label{eqpermu}
  \chi\left(\frac{\mu}{\mu-1}\right)-\chi(1)+\frac{1}{2\lambda_c}-\frac{1}{2\lambda_\Gamma(3+\eta)}=0\,,
\end{equation}
from which one gets $\mu\simeq 17.21$ (hence $x_{res}\sim 1.1\sim
{\cal O }(1)$). Fig. \ref{fig} shows the values of the parameter
$\mu$ for varying $\eta$. As one can see, $\mu$ approximates to
17.21 as $\eta\lesssim 10^{-7}$.

From the above results, one infers
 \[
 T(r_{res})\sim 1.2\, T_c\,, \quad  r_{res}\sim 0.9 \, r_c\,, \quad  R_s\sim 1.4
 \,r_c\,.
 \]

%\section{Conclusion}

Let us now discuss the obtained results in relation to the
velocity distribution of pulsars. At the moment, the statistical
analysis of pulsar population neither support nor rule out any
model/mechanism proposed to explain pulsar kicks. This is
essentially due to the lacking of correlation between neutron star
velocity and the other properties of neutron stars (see for
example \cite{lai} for a general discussion and references
therein).

Concerning the coefficients for the Lorentz violation, it is
convenient to adopt a standard inertial frame with respect to
which experimental measurements of these coefficients are reported
\cite{SunFrame}. In this frame the coefficients ${\bf a}_L$ and
${\bf c}_L$ (hence $\delta {\bf a}_L$ and $\delta {\bf c}_L$)
point along all possible orientation\footnote{
%As an example, one can have that ${\bf a}_L$
%and ${\bf c}_L$ may point along orthogonal directions: ${\bf a}_L$
%points along the north or south direction and ${\bf c}_L$ lies on
%the equatorial plane. In this particular case the effects of the
%Lorentz and CPT invariance violation on pulsar kicks may be
%relevant both for neutrinos propagating along $\Theta=\phi=0$
%(${\bf p}$ orthogonal to the equatorial plane and $\phi$ is the
%angle between ${\bf p}$ and ${\bf a}_L$) and $\Theta=\phi=\pi/2$
%(${\bf p}$ lies on the equatorial plane and $\phi$ is the angle
%between ${\bf p}$ and ${\bf c}_L$).
In particular, in the papers \cite{mewesneutrino,KostLSND} it has
been studied the general theory with $a_L^\mu$ and $c_L^{\mu\nu}$
which include all orientations, whereas in \cite{KostPRD70} the
$a_L^\mu$ coefficients point along the north direction, and
$c_L^{\mu\nu}$ are isotropic.}. Usually, it is assumed that
Lorentz violating parameters are independent on position
\cite{mewesneutrino,KostLSND,KostPRD70}. Nevertheless, as
discussed by Kosteleck\'y, Lehnert, and Perry in \cite{bertolami},
they may also exhibit a space-dependence, such that they may take
{\it arbitrary} different values for pulsars at different places.

\vspace{0.2in}

These considerations entail two basic features:

\begin{itemize}
  \item A priori there is no favored direction for the Lorentz violating
  coefficients. This results in a uniform distribution of pulsars, in agreement
with statistical analysis performed in Refs.
\cite{arzoumanian,lai,possenti}.
  \item The mechanism proposed in this paper does not require any
correlation between the velocity $V$ and other physical parameters
of pulsars, such as for example, the magnetic field $B$. The
reason is essentially due to the fact that in the Standard Model
Extension, the coefficients $a_L^\mu$ and $c_L^{\mu\nu}$ describe
the most general renormalizable effects that are possible in the
gauge-invariant neutrino sector, no matter what are their origin
at the Planck scale (they can be regarded as vacuum expectation
values of tensor operators arising from the spontaneous breaking
mechanism). As a consequence, a correlation among the coefficients
$a_L^\mu$ and $c_L^{\mu\nu}$ and the parameters characterizing the
pulsar properties, in particular the magnetic field, is not
necessarily expected. Such a result, indeed, seems to be
corroborated by recent analysis of Refs.
\cite{arzoumanian,lai,possenti} in which it is pointed out the
apparent lack of evidence in favor of the $B-V$
correlation\footnote{It should be noted that the analysis of Refs.
\cite{arzoumanian,lai,possenti} does not imply that the
Kusenko-Segr\'e mechanism for generate the pulsar kicks, as well
as those mechanisms that also rely on magnetic fields, do not
work. As pointed out by Kusenko in \cite{alex}, such mechanisms do
not predict a $B-V$ correlation since the relevant magnetic field
for the kick velocity is the magnetic field {\it inside} the hot
neutron star during the first seconds after the Supernova
collapse. Astronomical observations allow to infer the surface
magnetic field some millions of years later. These two fields are
not trivially correlated each other because of the complex
evolution of the magnetic field during the cooling process of
neutron star, which lead to a final (surface) magnetic field whose
configuration is different from the initial one, i.e. few seconds
after the onset of the Supernova (see \cite{alex} for a detailed
discussion).}.
\end{itemize}

%A simple choice is $\Theta=\phi$, i.e. $\delta {\bf c}_L$ and
%$\delta {\bf a}_L$ point along the north direction. With respect
%to the standard Sun-centered frame, it follows that the effects of
%the Lorentz invariance violation on pulsar kicks is maximal at
%$\Theta=0$, i.e. for high energy neutrinos which propagate
%parallel to celestial north or south. The effects appear sensibly
%reduced for the propagation in the equatorial plane determined by $\Theta=\pi/2$.

%In deriving (\ref{deltaGamma}) we have assumed that $\delta {\bf
%c}$ and $\delta {\bf a}$ are parallel, but in general one can have
%two preferred directions, hence two different angles (the angles
%between ${\bf {\hat p}}$ and $\delta {\bf c}$, and ${\bf {\hat
%p}}$ and $\delta {\bf a}$).

In conclusion, the origin of pulsar velocities has been studied in
the framework of Standard Model Extension. The effective
Hamiltonian for the time evolution of neutrinos (see Eq.
(\ref{heff})) gives rise to a term of the form $\sim \delta {\bbox
\Pi}\cdot {\bf p}$. The relative orientation of neutrino momenta
with respect to $\delta {\bbox \Pi}$ leads to the fractional
asymmetry necessary to  the generate the observed pulsar motion.
Future observations and statistical analysis on the velocity
distribution of pulsars might provide a scenario in which the
Lorentz and CPT invariance violation, as suggested by
Kosteleck\'y, might be tested on astrophysical scales.

%Certainly future experiments and observations on pulsar properties
%will allow a more deep understanding on the origin of pulsar
%kicks. This will also provide a further scenario in which the
%Lorentz invariance violation, as suggested by Kosteleck\'y theory,
%might be tested on astrophysical scales.

\acknowledgments It is a pleasure to thank V. Alan Kosteleck\'y
for comments and suggestions which have improved the paper. Thanks
also to A. Possenti for discussions on pulsar physics  and for
drawing to my attention to Refs. \cite{possenti}.
% which have led the paper to the final version.
%Thanks also to G. Raffelt
%for elucidations concerning some topics related to pulsar properties.
Research supported by PRIN 2004.

\vspace{0.1in}

\appendix

\section*{The Polytropic Model}

\vspace{0.1in}

The aim of this Appendix is to recall the main topics of the
polytropic model which describes the inner core of a protostar. To
this end, we shall refer to the paper by Barkovich, D'Olivo,
Montemayor and Zanella \cite{zanella}. The relevant equations for
the description of an isotropic neutrinosphere are the equation
for the hydrodynamical equilibrium
\begin{equation}\label{P(r)A}
  \frac{dP(r)}{dr}=-\rho(r)\, \frac{GM(r)}{r^2}\,,
\end{equation}
where $M(r)=4\pi \int_0^r dr^\prime r^{\prime\, 2}
\rho_T(r^\prime)$ ($\rho_T$ is the total density of matter), the
equation for the energy transport
\begin{equation}\label{F(r)A}
 F(r)=-\frac{1}{36} \frac{1}{\kappa \rho(r)}\frac{dT^2}{dr}\,,
\end{equation}
and the equation for the flux conservation
\begin{equation}\label{LuminosityA}
  F(r)=\frac{L_c}{4\pi r^2}\,,
\end{equation}
where $L_c$ is the luminosity of the protostar. The equation of
state with adiabatic index $\Gamma$ is \cite{shapiro,zanella}
\begin{equation}\label{PA}
  P(r)=K\rho^\Gamma\,,\quad K=\frac{T_c}{m_n\rho_c^{1/3}}
\end{equation}
where $T_c$ and $\rho_c$ are the temperature and the matter
density of the core. From Eqs. (\ref{PA}) and (\ref{P(r)A}) one
gets
\begin{equation}\label{drho}
  \frac{d\rho^{\Gamma-1}}{dr}=-\frac{\lambda_\Gamma r_c\rho_c^{\Gamma-1}}{M_c}
  \frac{M(r)}{r^2}\,,
\end{equation}
where $M_c$ is the core mass. A function which represents an
extremely good approximation for $\rho$ is given by \cite{zanella}
\begin{equation}\label{approxrho}
  \rho^{\Gamma-1}(r)=\rho^{\Gamma-1}_c \left[
  \lambda_\Gamma \left(\frac{r_c}{r}-1\right)m(r)+1\right]\,,
\end{equation}
where $m(r)=\mu+(1-\mu)r_c/r$. Notice that $m(r_c)=1$. The
parameter $\mu$ is determined by using the equation $\rho(R_s)=0$
($R_s$ is the radius of the star), i.e.
\begin{equation}\label{muA}
  \mu=\left[\frac{R_s}{\lambda_\Gamma(R_s-r_c)}-\frac{r_c}{R_s}\right]\frac{R_s}{R_s-r_c}\,.
\end{equation}
Eq. (\ref{approxrho}) can be easily recast in the form
\begin{equation}\label{rhoA}
  \rho^{\Gamma-1}(r)=\rho_c^{\Gamma-1}\left[
  a'\left(\frac{r_c}{r}\right)^2+b'\frac{r_c}{r}+c'\right]\,,
\end{equation}
where the coefficients $a', b', c'$ are defined in (\ref{abc}).
Eqs. (\ref{LuminosityA}) and (\ref{F(r)A}) allow to infer the
temperature profile
\begin{equation}\label{T2A}
  \frac{dT^2}{dr}=-\frac{9\kappa L_c}{\pi r^2}\rho\,,
\end{equation}
with $\rho$ given by (\ref{rhoA}). For $\Gamma=4/3$ the solution
of the integro-differential equation (\ref{T2A}) is
\begin{equation}\label{T(r)A}
  T(r)=T_c\sqrt{2\lambda_c[\chi(r_c/r)-\chi(1)]+1}\,,
\end{equation}
with $\chi(r_c/r)$ defined in (\ref{chi}).

\end{document}